\documentclass[11pt,a4paper]{article}
\usepackage[utf8]{inputenc}
\usepackage{tikz}
\usepackage{graphicx}
\usepackage{subcaption}
\usepackage[T1]{fontenc}
\usepackage{authblk}
\usepackage{amsmath}
\usepackage{url}

\usepackage[a4paper, total={6in, 8in}]{geometry}

\title{Using Causal Inference to Explore Government Policy Impact on Computer Usage
}

\author[1]{Mingjia Zhu \footnote{Both authors contributed equally to the work.}}
\author[2]{Lechuan Wang $^*$}
\author[3]{Julien Sebot}
\author[3]{Bijan Arbab}
\author[1, 4]{Babak Salimi}
\author[2,4]{Alexander Cloninger}
\affil[1]{University of California San Diego, Computer Science and Engineering}
\affil[2]{University of California San Diego, Department of Mathematics}
\affil[3]{Intel Corporation}
\affil[4]{University of California San Diego, Halicio{\u g}lu Data Science Institute}

\usepackage{pgfplots}

\begin{document}
\maketitle

\begin{abstract}
We explore the causal relationship between COVID-19 lockdown policies and changes in personal computer usage.  In particular, we examine how lockdown policies affected average daily computer usage, as well as how it affected usage patterns of different groups of users. This is done through a merging of the Oxford Policy public data set, which describes the timeline of implementation of COVID policies across the world, and a collection of Intel's Data Collection \& Analytics (DCA) telemetry data, which includes millions of computer usage records and updates daily.
Through difference-in-difference, synthetic control, and change-point detection algorithms, we identify causal links between 
the increase in intensity (watts) and time (hours) of computer usage and the implementation of work from home policy. We also show an interesting trend in the individual’s computer usage affected by the policy. We also conclude that computer usage behaviors are much less predictable during reduction in COVID lockdown policies than during increases in COVID lockdown policies. 
\end{abstract}

%
%

\section{Introduction}

During the COVID-19 pandemic, most governments issued policies to control the spread. For example, some governments prohibit public activities and close school/onsite working opportunities. In addition, some ask their citizens to get vaccines and constrain the price of essential products which are important to people’s lives. As a result, people have changed their living habits, including their work, school, and leisure. For work, an increasing number of companies allow their employees to work from home, which has some influence on the daily usage of computers. 

We investigate the relationship between COVID policies and the daily usage of computers.  Successful analysis of this effect can help the industry change their computer structures accordingly, provide a better customer experience, and own a bigger market. We mainly explore how many hours and watts the average computer usage increases because of the workplace closing policy. In addition, we move forward with individual users to obtain a more fine-grained result.

We hypothesized prior to analysis that there is a significant and causal effect from workplace closure on computer usage and behavior.  Additionally, we hypothesized that “lighter and more mobile” devices, and user facing devices, are the one whose usage increased most.  There exists a growing body of literature that has observed correlations between COVID lockdown and these types of effects, including an increase in WiFi usage on smartphones \cite{li2021impact}, surveys indicating increased computer usage \cite{bahkir2020impact, lee2021information}, and participation studies evaluating the health effects of such increase usage \cite{tyagi2021effects}.  However, all studies of this hypothesis, to our knowledge, examine only correlational and observational effects of COVID lockdown policies.  

To identify the causal functionality of those policies, we implement the difference in the difference model and the synthetic control method. In addition, we make some hypotheses about confounders and do experiments to prove our hypotheses.
Based on our investigation, the activation of the workplace closing policy caused the daily hours and intensity of computer usage to increase for notebooks, desktops, and 2-in-1s with CPU i5, i7, and i9. The effect of the policy deactivation is much slower and weaker than that of activation. We also use the change points detection algorithm to discover that the individual’s computer usage is correlated with lockdown patterns.

\section{Method}
\subsection{Datasets}
\subsubsection{Oxford Covid-19 Government Response Tracker}
The Oxford Covid-19 Government Response Tracker (OxCGRT) \cite{hale2021global} includes policies in response to the COVID-19 pandemic issued by governments all over the world, including governments of countries, regions, and states. It contains 23 policy indicators of government response, including containment and closure policy, economic policy, health system policy, and vaccination policy. Ordinal scales were used in most indicators' records to represent the policy's strictness level.

\subsubsection{Intel DCA Dataset}

Millions of people come to Intel web site per month looking to update their device drivers for wifi, Bluetooth, graphics, or need to update their BIOS, or Firmware. Intel Driver \& Support Assistance (IDSA) web site provides an automated software program that examines each device to discover exact versions of CPU, motherboard, graphics cards, network interfaces, etc. as well as the associated driver, bios, and firmware versions.  IDSA then compares each device’s discovered HW and SW version against an internal DB to see if updated version are available and upon user demand updates the device to the latest versions.  During this process each user is asked to explicitly Opt-In or Opt-Out of sending device usage telemetry data back to Intel with a strict EULA limiting the use of the data to improving performance, and design.

Approximately 40\% of people Opt-In, while 60\% of people Opt-Out of sending telemetry data back to Intel.  The data from these millions of worldwide devices from is the primary source of data used in this study. Over 1,400 device specific parameter are collected from each device (CPU power used, temperature, network traffic, HDD read/write traffic, applications launch times and  durations used, frame per second experiences for each window, hard and soft page faults, etc.  There are no personal information of any kind collected from each device, e.g. device serial number, IP or MAC addressed, specific URLs visited, email, or anything that can be used to track device specific data back to individual users.  The main purpose of this telemetry effort is to study and understand user experiences from millions of devices as used on a daily basis \cite{su2023product}.

For that purpose, a device is assigned a Global Unique Identifies (GUID), which allows device usage data over time to be associated with the same device without compromising user’s identity. 
The GUID can be used to stitch data points gathered over time to the same device.  GUIDs allow us to observe same device behavior over time to perform various analytics, yet preserver user level PII (Private Identifiable Information) integrity.

Intel DCA Dataset includes variables about computer usage. These variables are statistics from records reported by computer users, which reflect computer usage from quantity and intensity aspects. The dataset also contains the computer type of each record. Below are some variables we have used in the dataset:
\begin{itemize}
\item Date: the report date
\item GUID: Global Unique Identifies.  This allows device usage data over time to be associated with the same device without compromising user’s identity. 
\item Computer type 
\begin{itemize}
    \item Notebook
    \item Desktop
    \item 2-in-1 : convertible tablet-notebook computers
    \item NUC: very small form factor battery-less computer, tends to be providing a 24/7 service like disk server or public display.
\end{itemize}

\item Processor Line: organized by target form factors (U thin and light mobile, H bigger screen/ thicker notebooks)
\item CPU family : How fast/high end cpus are (from low end Pentium/Celeron, to i3,5,i7,i9 highest end)
\item Vpro Enabled: proxy indicator of whether the system part of a fleet (work/school computer).  
Vpro is a paid feature that provides fleet management capabilities, so it is expected to be absent from most personal computers and be present only on some of the professional/school laptops (the ones for larger organizations). 
\item C0 state time: average hours system is in use, aka C0 state according to Advanced Configuration and Power Interface
\end{itemize}

We merge these two data sets to analyze the change of computer usage at the country level. Our study utilizes around 50 countries worldwide and focuses on the data in the United States and China. 
The main reason for this choice is the time lag in implementation and deactivation of policies, as both lock down and relaxation of lock down were implemented approximately two months earlier in China than in the US.
This creates a natural experiment condition for the causal inference. 
Another reason for this choice is the large number of computers available in both countries, as well as a diversity of computer uses in each country.  
Furthermore, we do cover other countries which have meaningful data as well. We classify computers based on their chassis types, processor line, and CPU family in our analysis. This enables us to analyze how different groups of people change their habits of using computers.

Among various policies, we choose to explore the effect of the C2 - Workplace Closing Policy indicator on computer usage, as we hypothesize that work from home policies and unemployment were a leading cause of the changes in the length of time people use computers. The codings for workplace closing policy are in Table \ref{tab:oxford}.

\begin{table}[]
\centering
\resizebox{\columnwidth}{!}{%
\begin{tabular}{|l|l|}
\hline
0 & No measures                                                                                          \\ \hline
1 & Recommend closing (or recommend work from home) or all businesses open with alterations resulting in significant differences compared to non-Covid-19 operation \\ \hline
2 & Require closing (or work from home) for some sectors or categories of workers                        \\ \hline
3 & Require closing (or work from home) for all-but-essential workplaces (e.g., grocery stores, doctors) \\ \hline
\end{tabular}%
}
\caption{Oxford Government Response Tracker Codes for Workplace Closure}\label{tab:oxford}
\end{table}

We mainly investigate the effect of activation (change from 0 to 3) of the policy and deactivation (change from 3 to 2). For activation, we consider 0 as “no treatment” and 3 as “treatment”; for deactivation, we consider 3 as “no treatment” and 2 as “treatment.”  Unfortunately using 0 as a deactivation treatment is difficult as many regions did not reduce their policies to attain a score of 0 until several years after the pandemic began, and well after behavior in response to the pandemic had shifted. 

\section{Methods}


\subsection{Difference-in-differences}

We employ the Difference-in-Differences (DiD) method~\cite{o2016estimating,villa2016diff}, a widely adopted evaluation approach for deriving causal inferences from observational panel data. This method compares changes in outcomes over time between a treated population and a control population, providing valuable insights into the impact of the program under investigation.

The operational framework of the DiD method can be elucidated through the subsequent stages (where A, B, and C are referred to in the accompanying graph): 
\begin{enumerate}
    \item Determine the difference in outcomes between the treatment and control groups prior to intervention (A).
    \item Ascertain the difference in outcomes between the treatment and control groups subsequent to intervention (B).
    \item Compute the difference between pre-treatment and post-treatment differences as {\em average treatment effect}, represented by  C = B - A.
\end{enumerate}

The validity of DiD critically relies on specific assumptions, including: (1) Intervention is unrelated to outcome at baseline, ensuring that the allocation of intervention was not determined by the outcome. (2) Treatment and control groups must exhibit parallel trends in outcomes. (3) The composition of intervention and comparison groups remains consistent for repeated cross-sectional design. (4) Absence of spillover effects among groups (Columbia Public Health). Among these, assumption (2) holds paramount importance in DiD, as any violation of the parallel trend assumption could introduce bias into the causal effect estimation~\cite{bertrand2004much}. 

DID is usually implemented as an interaction term between time and treatment group dummy variables in a regression model as follows:
\begin{align*}
    Y_{i,t} = \alpha + \beta_0 D_i(t) + \beta_1 X_i + \gamma t + \epsilon_i(t)
\end{align*}
    
Here, \( \alpha \) embodies the main effect, \( \beta_0 \) symbolizes the causal effect, \( \beta_1 \) serves as the intercept, \( t \) denotes the time step, and \( \epsilon_i(t) \) accounts for the noise or unobservable error terms. This formulation effectively captures the complex interplay between variables, allowing for a robust estimation of the program's impact.

In the context of the present project, two DiD experiments were conducted. Initially, a comparative study was made using Taiwan as the control group and California as the treatment group, the rationale being Taiwan's non-implementation of COVID policies during the pandemic. Subsequently, a more extensive examination was carried out using California, Arizona, Taiwan, Denmark, Spain, and Chile as control groups, with China serving as the treatment group. This approach exploited China's early implementation of workplace closing policy as a natural experiment condition, distinguishing it from other global counterparts.

\subsubsection{Synthetic Control Method}
We also employ the synthetic control method~\cite{abadie2021using,abadie2015comparative}. This statistical technique is utilized to assess the causal impact of an intervention or treatment on either a single unit or a group of units within a comparative case study framework. The method constructs a synthetic unit by combining control groups with specific weights and subsequently compares this synthetic control to the treatment group. This approach facilitates the creation of a counterfactual unit that closely tracks the pre-intervention trends of the treatment group, thereby enabling a more robust inference regarding causality.

Consider a scenario where we have \(J+1\) units, with the first unit being the treatment group and the remaining \(j = 2, \dots, J+1\) units as control groups. Let \(T_0\) be the time at which the treatment occurs. For each unit \(j\) and time \(t\), we denote \(Y_{jt}^N\) as the outcome without treatment and \(Y_{jt}^I\) as the outcome with treatment. The goal of synthetic control is to estimate the counterfactual outcome \(Y_{1t}^N\) using the outcomes from the control groups, \(Y_{2t}^N, Y_{3t}^N, \dots, Y_{Jt}^N\), and compare it with the actual observed outcome \(Y_{1t}^I\). 

Formally, the synthetic control is defined as a weighted average of the control units. Given weights \(W = (w_2, w_3, \dots, w_{J+1})\), the synthetic control estimation of \(Y_{jt}^N\) is computed as:
\[
Y_{jt}^N = \sum_{j=2}^{J+1} w_j Y_{jt}^N
\]

The determination of weights can be viewed as a constrained linear regression problem. It requires the estimation of weights that minimize a loss function reflecting the pre-intervention difference between the synthetic control and the treatment group. To ensure the validity and reliability of the synthetic control, constraints are applied to confine all weights between 0 and 1, and the sum of weights is set to equal 1. This helps prevent interpolation that might lead to unrealistic or infeasible data. The weights are typically estimated using methods such as sequential least-square programming, aiming to minimize the discrepancy between the synthetic control and the treatment group before the intervention time.

The synthetic control method thus provides an elegant solution to causal inference challenges, particularly when dealing with heterogeneous treatment effects or when a clear-cut control group is unavailable. Its application complements traditional methods like difference-in-differences and offers additional robustness in empirical analysis.




This code will generate a graph where both the treatment and synthetic control lines have small random fluctuations, and the two curves are approximately similar before the intervention, as required by the synthetic control method.

\subsection{Offline Breakpoint Detection Methods}

Change-point detection algorithms are essential in analyzing non-stationary and time-series signals, as they aim to identify sudden alterations or "breakpoints" in behavior \cite{aminikhanghahi2017survey}. These shifts may correspond to transitions between different states or regimes in the underlying system. In this context, the simplest change-point algorithms aim to detect a given number of breakpoints that segment the signal into low-complexity parts, such as constant or linear pieces \cite{aminikhanghahi2017survey}.

We primarily focus on offline detection algorithms in this paper. Unlike online methods, which analyze the data in a causal or sequential manner, offline algorithms are supplied the entire time series at once. This often allows for more computationally intensive analysis and may result in more accurate detections.

In scenarios where the number of breakpoints is known and the goal is to decompose the signal into piecewise constant segments, the task becomes a discrete constrained optimization problem. The objective is to minimize the following loss function:
\begin{align*}
\sum_{i=1}^{K}\sum_{t\in I_i} |y_t - \overline{y_i}|^2,
\end{align*}
where ${I_i}_{i=1}^K$ denotes connected and disjoint time intervals, and $\overline{y_i}$ represents the mean value over the interval $I_i$.

When the number of changes is unknown, the objective function must be slightly modified to include a penalty term that accounts for the complexity of the segmentation. Criteria such as Akaike's Information Criterion (AIC) or the Bayesian Information Criterion (BIC) may be employed, as illustrated by:
\begin{align*}
\text{Objective} + \lambda \times (\text{AIC or BIC criteria}),
\end{align*}
where $\lambda$ controls the trade-off between fitting the data and penalizing complexity.

In our study, changepoint algorithms are deployed to detect the most substantial jumps in the data using an unsupervised method. We then scrutinize the identified breakpoints post hoc to ascertain if they coincide with major events, such as COVID lockdowns. Additionally, we assess the stability of the detected breakpoints by investigating their sensitivity to different hyperparameters. The results offer valuable insights into the dynamics of the studied system and demonstrate the efficacy of offline breakpoint detection algorithms.

\section{Results}

In this section, we detail the data analysis performed to investigate our hypotheses.  As a reminder, we hypothesize that there is a significant and causal effect from workplace closure on computer usage and behavior.  Additionally, we hypothesize that “lighter and more mobile” devices, and user facing devices, are the one whose usage increased most.  The devices in order of "mobility" would be 2 in 1 are tablet-laptops (small and light), then laptops, then desktops, then NUCs (these tend to be used as mini file servers or to drive screens in public spaces).  We also hypothesize that workplace closure had a significant effect on the behavior of individual users, with users transitioning to more gaming and communication applications.

\subsection{Activation of workplace closing policy on computer usage time}
\subsubsection{Difference in differences, without confounders}

\begin{figure}[htbp]
    \centering
    \includegraphics[width=0.5\textwidth]{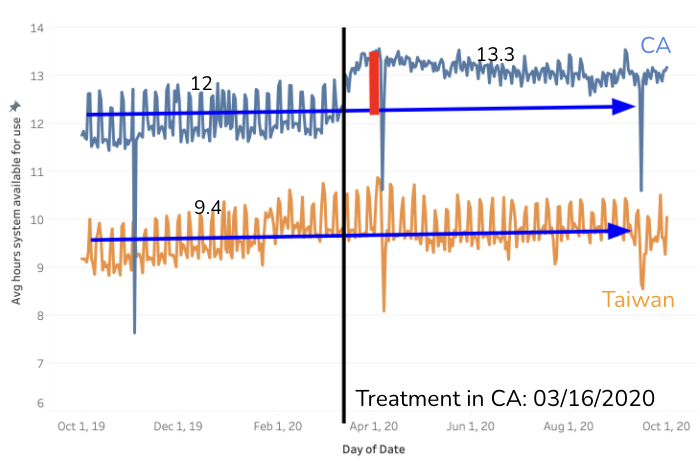}
\caption{Panel OLS analysis illustrating the effect of workplace closing policy on average computer usage. The estimated policy effect is positive, indicating an increase in computer usage post-policy implementation. The near-zero p-value suggests statistical significance.}
    \label{fig:mesh1}
\end{figure}

For the initial analysis, we employed a baseline Difference-in-Differences (DiD) model, without assuming any confounders, revealing a 1.3-hour increase in computer usage time before and after the policy change. These findings are illustrated in Figure~\ref{fig:mesh1}. Our study focused on notebooks equipped with U-processors (thin and light notebooks).  We used Taiwan as the control group and California as the treatment group.  Taiwan was chosen as a control because their effective implementation of quarantine allowed them to remain COVID free until Spring 2022, and therefore they didn't implement any stay at home orders or other lockdown procedures during the period studied.

In the graph, a black line marks the date when California implemented its workplace closing policy on March 16, 2020. A top blue arrow indicates the counterfactual trend of average computer usage, under the assumption that the pre-treatment trends between the two groups would have continued in parallel had there been no intervention. By comparing the actual trend in the treatment group with this counterfactual, we concluded that there was a significant increase in average computer usage following the policy's enactment. This change is highlighted by the bold red segment in Figure~\ref{fig:mesh1}, representing the treatment effect.


\begin{figure*}
\centering
\includegraphics[width=0.5\textwidth]{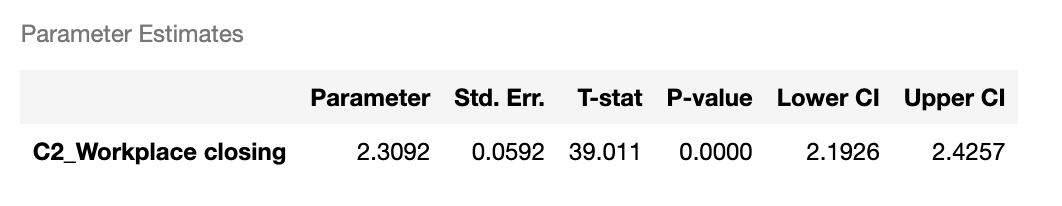}
\caption{Estimation of policy effect weight (2.3092), p-value, and confidence interval.}
\label{fig:parameterEstimation}
\end{figure*}

Therefore, we concluded that there exists a  dependence between the implementation of the workplace closing policy in California and the increase in average computer usage time. However, the estimated policy effect weight of 2.3092 is substantially greater than the actual observed increase of 1.3 hours. This notable discrepancy led us to investigate the impact of potential confounding variables. For a more detailed understanding of the magnitude of the policy effect, the estimated policy effect weight, p-value, and confidence interval are presented in Figure~\ref{fig:parameterEstimation}.

\begin{figure}[h]
\centering
\includegraphics[width=0.5\textwidth]{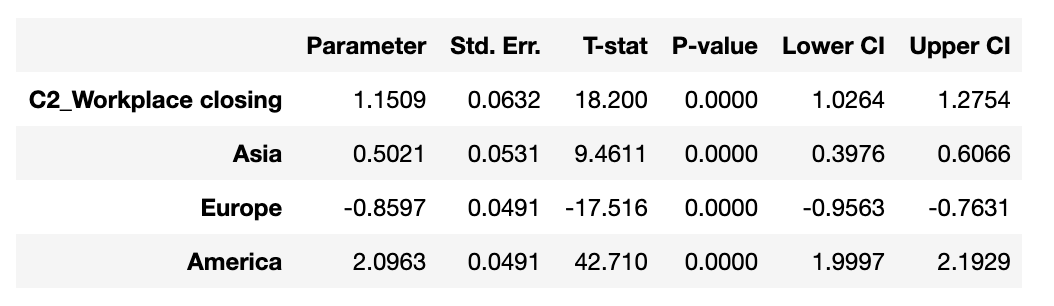}
\caption{Comparison of average computer usage across continents and the impact of workplace closing policy.}
\label{fig:continentComparison}
\end{figure}

\subsubsection{Continent information affects the baseline computer usage 
}
We observed that baseline computer usage is influenced by geographical location, specifically the country's continent. To corroborate this finding, we examined data from over 15 countries and identified a general trend: computer usage was lowest in Europe, medium in Asia, and highest in America. In this extended model, Taiwan serves as the control group, while seven other countries from these continents act as treatment groups.

Upon adjusting for continental differences, the estimated policy effect weight reduced to 1.15, aligning more closely with the observed increase of 1.3 hours. This suggests that the model, which accounts for continental variations, provides a more accurate estimate of the policy's impact. However, it's important to note that several countries deviated from this general trend. Additionally, the lack of a suitable control group for some countries limited the model's effectiveness.  See Table in Figure~\ref{fig:continentComparison} for details of the analysis.
\subsubsection{Effects of the Workplace Closing Policy on Different CPUs and Computer Types}

In our analysis, we observed that the type of computer used significantly influenced baseline computer usage, necessitating their inclusion as control variables in our models. Specifically, 2-in-1 computers exhibited the lowest baseline usage, followed by notebooks, and finally, desktops registering the highest usage. To facilitate a more nuanced analysis, we disaggregated the dataset based on both computer and CPU types.

To generate robust comparisons, we employed the synthetic control method to construct a suitable control group. China served as the primary treatment group due to its early implementation of workplace closure policies on January 26, 2020. The synthetic control groups consisted of data from over 25 countries, including the United States.

\begin{figure}[htbp]
    \centering
    \begin{minipage}{.3\textwidth}
        \begin{subfigure}{\textwidth}
            \centering
            \includegraphics[width=\textwidth]{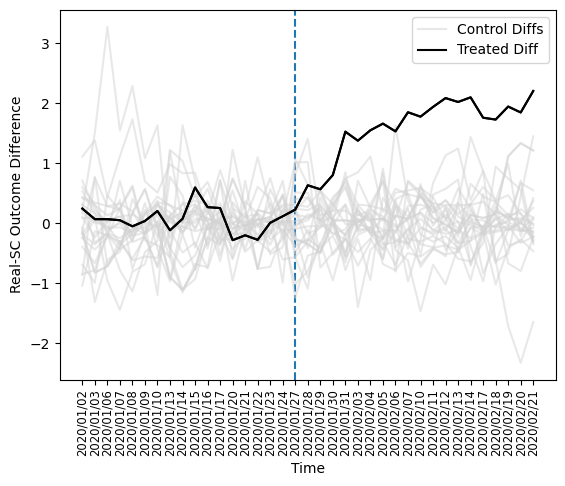}
            \caption{Treatment Difference for 2-in-1 Computers}
        \end{subfigure}\\
        \begin{subfigure}{\textwidth}
            \centering
            \includegraphics[width=\textwidth]{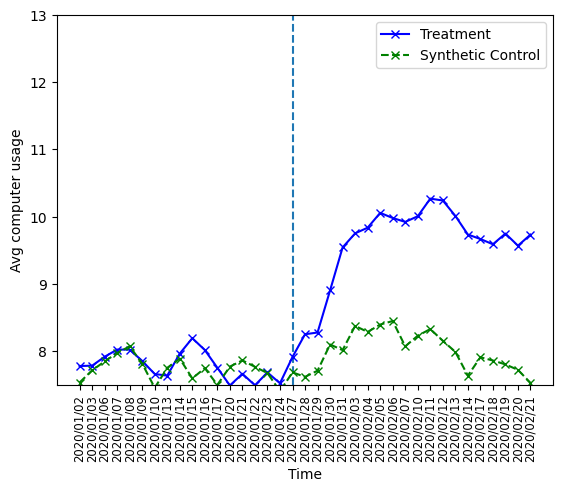}
            \caption{Treatment Difference for 2-in-1 Computers}
        \end{subfigure}    
    \end{minipage}
    \hfill
    \begin{minipage}{.3\textwidth}
        \begin{subfigure}{\textwidth}
            \centering
            \includegraphics[width=\textwidth]{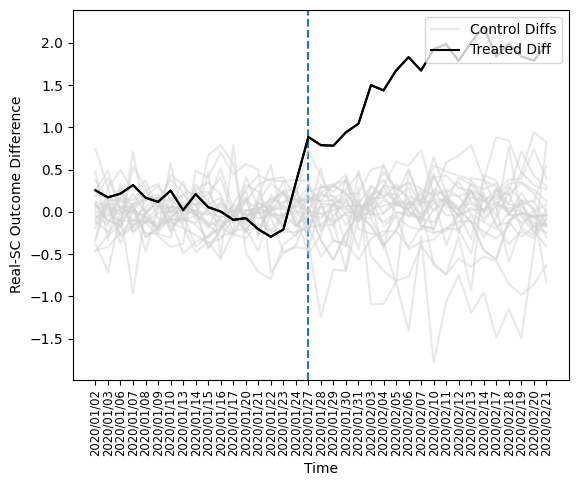}
            \caption{Treatment Difference for Notebooks}
        \end{subfigure}\\
        \begin{subfigure}{\textwidth}
            \centering
            \includegraphics[width=\textwidth]{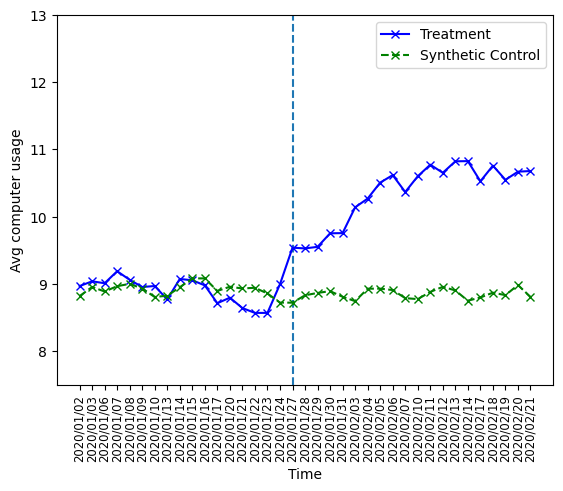}
            \caption{Treatment Difference for Notebooks}
        \end{subfigure}    
    \end{minipage}
    \hfill
    \begin{minipage}{.3\textwidth}
        \begin{subfigure}{\textwidth}
            \centering
            \includegraphics[width=\textwidth]{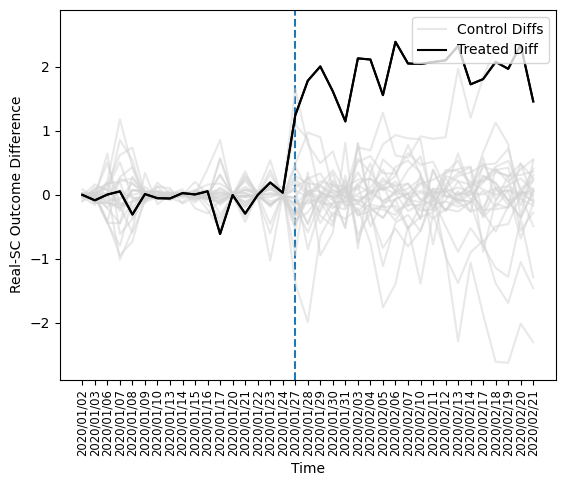}
            \caption{Treatment Difference for Desktops}
        \end{subfigure}\\
        \begin{subfigure}{\textwidth}
            \centering
            \includegraphics[width=\textwidth]{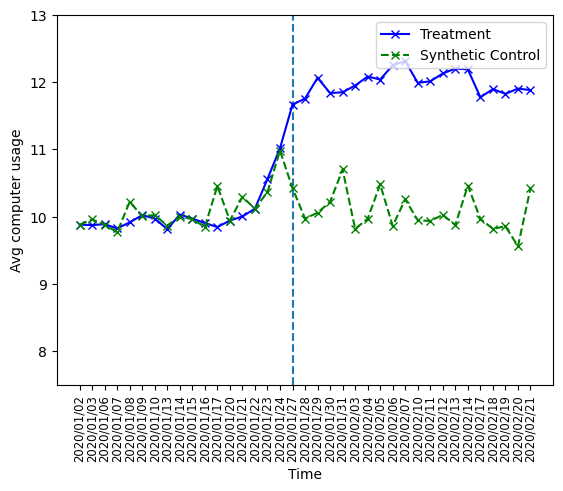}
            \caption{Treatment Difference for Desktops}
        \end{subfigure}    
    \end{minipage}
    \caption{Comparison of Policy Effects Across Various Computer Types}
    \label{fig:allComputers}
\end{figure}

As depicted in Figure~\ref{fig:allComputers}, the treatment effects are clearly visible. The blue dashed lines indicate the date of treatment implementation. In the first set of graphs, the black lines represent the treatment difference, which is notably higher than the light grey lines showing control differences. These outcomes were found to be statistically significant with p-values below 0.05. The bottom set of graphs in Figure~\ref{fig:allComputers} further confirm these findings, as they display a significant post-treatment increase in computer usage in the treatment group as compared to synthetic controls. It should be noted that these graphs focus on i7 CPUs (higher end processors); however, a similar trend was observed across different CPU types.

\begin{table}[htbp]
\centering

\resizebox{\columnwidth}{!}{%
\begin{tabular}{|c|cc|cc|cc|}
\hline
\multicolumn{1}{|l|}{} & \multicolumn{2}{c|}{I5}           & \multicolumn{2}{c|}{I7}           & \multicolumn{2}{c|}{I9}          \\ \hline
\multicolumn{1}{|l|}{} &
  \multicolumn{1}{c|}{\begin{tabular}[c]{@{}c@{}}Increase in \\ computer usage (hrs)\end{tabular}} &
  Avg system count &
  \multicolumn{1}{c|}{\begin{tabular}[c]{@{}c@{}}Increase in \\ computer usage (hrs)\end{tabular}} &
  Avg system count &
  \multicolumn{1}{c|}{\begin{tabular}[c]{@{}c@{}}Increase in \\ computer usage (hrs)\end{tabular}} &
  Avg system count \\ \hline
2-in-1                 & \multicolumn{1}{c|}{2.2} & 40935  & \multicolumn{1}{c|}{2.3} & 46029  & \multicolumn{1}{c|}{N/A} & 91    \\ \hline
Notebook               & \multicolumn{1}{c|}{1.7} & 385454 & \multicolumn{1}{c|}{1.8} & 359725 & \multicolumn{1}{c|}{2.6} & 2598  \\ \hline
Desktop                & \multicolumn{1}{c|}{1.4} & 236183 & \multicolumn{1}{c|}{1.5} & 173033 & \multicolumn{1}{c|}{1.8} & 14878 \\ \hline
\end{tabular}%
}
\caption{Increase in Computer Usage and System Count for Different CPUs and Computer Types During Workplace Closure Activation Timeframe}
\label{tab:cpu_usage_increase}
\end{table}

\subsubsection{Final Results with Vpro Percentage as a Confounder}

To further improve the model, we incorporated another variable: the Vpro percentage, which represents the proportion of computers equipped with Vpro technology in each country or region. Vpro, a paying feature focused on fleet management and security for larger organizations, is typically found in work computers.  We hypothesized that work and personal computers would exhibit different usage patterns, particularly as remote work becomes more prevalent. Our dataset only includes Vpro-enabled data for the i5, i7, and i9 CPUs; thus, we maintain the focus on i7 CPUs (higher end processors) across all computer types for consistency.

Our analysis yielded three primary findings, which are summarized in Table~\ref{tab:cpu_usage_increase}. First, we observed a statistically significant increase in average computer usage across all types of computers, albeit with variations. Specifically, 2-in-1 computers showed the most substantial increase in usage, while desktop computers showed the least. This trend aligns with the initial baseline usage rates for these types of computers: 2-in-1 computers started with lower baseline usage, whereas desktops had higher baseline usage. Lastly, after adjusting for the Vpro percentage as a confounding factor, the p-values from randomization inference tests approached zero for all three types of computers. This result provides strong evidence that the observed changes in computer usage were a rapid response to the policy implementation.


\begin{figure}[htbp]
    \centering
    \begin{minipage}{.3\textwidth}
        \begin{subfigure}{\textwidth}
            \centering
            \includegraphics[width=\textwidth]{figs/image41.png}
            \caption{Deactivation Effect for 2-in-1 Computers}
        \end{subfigure}\\
        \begin{subfigure}{\textwidth}
            \centering
            \includegraphics[width=\textwidth]{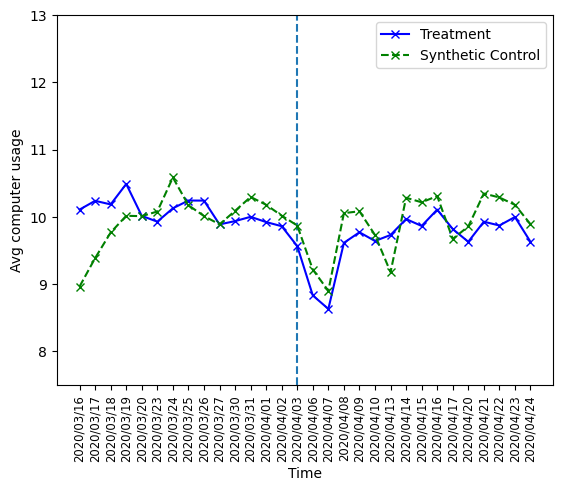}
            \caption{Deactivation Effect for 2-in-1 Computers}
        \end{subfigure}    
    \end{minipage}
    \hfill
    \begin{minipage}{.3\textwidth}
        \begin{subfigure}{\textwidth}
            \centering
            \includegraphics[width=\textwidth]{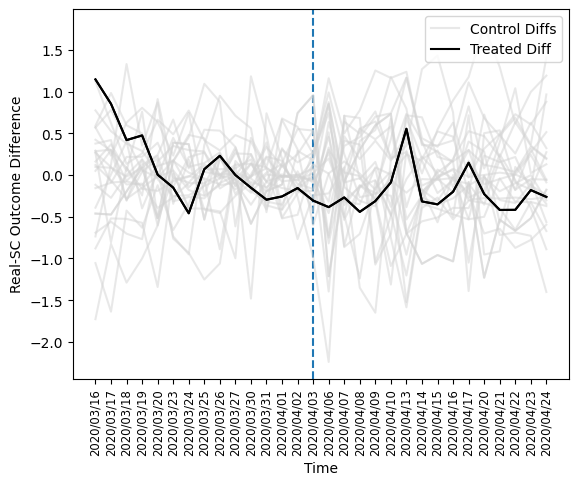}
        \end{subfigure}\\
        \begin{subfigure}{\textwidth}
            \centering
            \includegraphics[width=\textwidth]{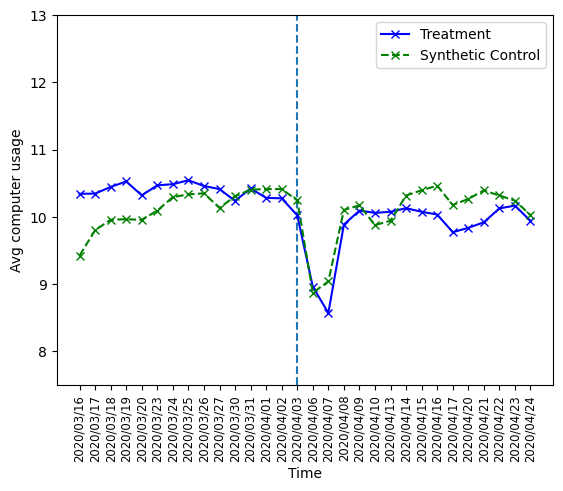}
            \caption{Deactivation Effect for Notebooks}
        \end{subfigure}    
    \end{minipage}
    \hfill
    \begin{minipage}{.3\textwidth}
        \begin{subfigure}{\textwidth}
            \centering
            \includegraphics[width=\textwidth]{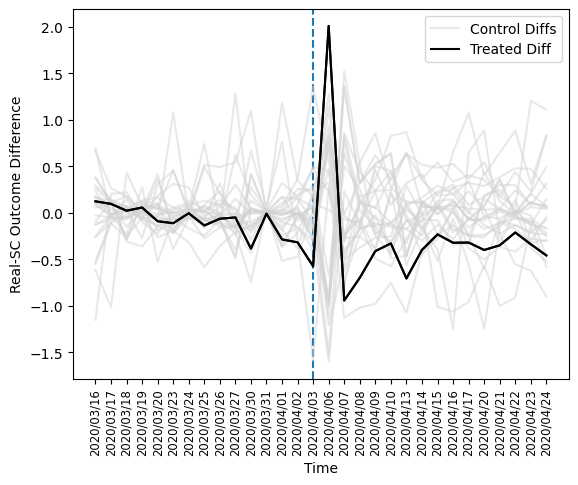}
            \caption{Deactivation Effect for Desktops}
        \end{subfigure}\\
        \begin{subfigure}{\textwidth}
            \centering
            \includegraphics[width=\textwidth]{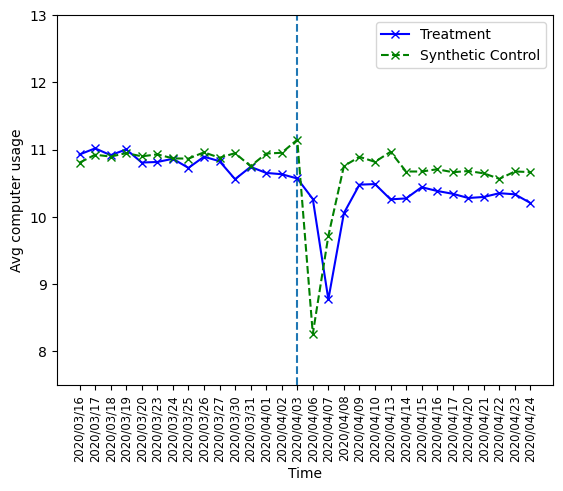}
            \caption{Deactivation Effect for Desktops}
        \end{subfigure}    
    \end{minipage}
    \caption{Comparison of Deactivation Effects Across Various Computer Types}
    \label{fig:deactivationEffect}
\end{figure}

\subsection{Deactivation of workplace closing policy on computer usage time}
We applied the same analysis model to the period following the deactivation of workplace closing policies. The term "deactivation" in this context refers to the easing of initial policies, specifically from requiring the closure of "all-but-essential workplaces" to a less stringent requirement affecting only "some sectors or categories of workers." China serves as the treatment group because it was the first to deactivate these policies on April 2, 2020, providing an ideal comparison group.  The raw numbers across various computer types can be viewed in Table \ref{deactivation}.

The synthetic control method was used again, employing the same control groups but with adjusted weights to better match the treatment group during the pre-treatment period. The findings are illustrated in Figure~\ref{fig:deactivationEffect}. Note that the results contain spikes due to data reporting issues, making it challenging to draw conclusions from these specific data points.

In general, computer usage within the treatment group was either slightly lower than or similar to that of the synthetic controls. The differences between them were either negligible or approached zero. The overall average computer usage exhibited only a minor decrease, and the extent of this decrease during the deactivation period was significantly less than the increase observed during the activation of workplace closing policies. The p-values obtained from randomization inference were not statistically significant across all three types of computers, suggesting that the changes in usage were likely random. These findings indicate that while the activation of workplace policies had a pronounced impact on computer usage, the deactivation of these policies did not lead to immediate or strong changes.

\begin{table}[]

\centering
\resizebox{\columnwidth}{!}{%
\begin{tabular}{|c|cc|cc|cc|}
\hline
\multicolumn{1}{|l|}{} & \multicolumn{2}{c|}{I5}           & \multicolumn{2}{c|}{I7}            & \multicolumn{2}{c|}{I9}          \\ \hline
\multicolumn{1}{|l|}{} &
  \multicolumn{1}{c|}{\begin{tabular}[c]{@{}c@{}}Decrease in \\ computer usage (hrs)\end{tabular}} &
  Avg system count &
  \multicolumn{1}{c|}{\begin{tabular}[c]{@{}c@{}}Decrease in \\ computer usage (hrs)\end{tabular}} &
  Avg system count &
  \multicolumn{1}{c|}{\begin{tabular}[c]{@{}c@{}}Decrease in \\ computer usage (hrs)\end{tabular}} &
  Avg system count \\ \hline
2-in-1                 & \multicolumn{1}{c|}{0.2} & 59421  & \multicolumn{1}{c|}{0.25} & 65236  & \multicolumn{1}{c|}{N/A} & 104   \\ \hline
Notebook               & \multicolumn{1}{c|}{0.4} & 521839 & \multicolumn{1}{c|}{0.1}  & 488788 & \multicolumn{1}{c|}{N/A} & 3782  \\ \hline
Desktop                & \multicolumn{1}{c|}{0.2} & 287623 & \multicolumn{1}{c|}{0.5}  & 215520 & \multicolumn{1}{c|}{0.1} & 20190 \\ \hline
\end{tabular}%
}
\caption{Decrease in Computer Usage and System Count During Workplace Closure Deactivation Timeframe}\label{deactivation}
\end{table}

\subsubsection{Other Confounders}
In our analysis, we also considered additional potential confounders such as Gross Domestic Product (GDP), unemployment rate, Google mobility data, and the demographic percentage of the population aged between 15 and 55 years. We hypothesized that a higher GDP and lower unemployment rate could indicate a greater proportion of white-collar workers, thus potentially increasing computer usage. Similarly, a younger population demographic and higher Google mobility data, which represents time spent at home or work, could also influence computer usage rates. However, despite the theoretical rationale for including these variables, they neither significantly improved the fit of our synthetic control model nor materially impacted the study's conclusions.



\subsection{Change point analysis}
\subsubsection{Activation of Workplace Closing Policy on Intensity of Usage}

\begin{figure}[htbp]
    \centering
    \includegraphics[width=0.7\textwidth]{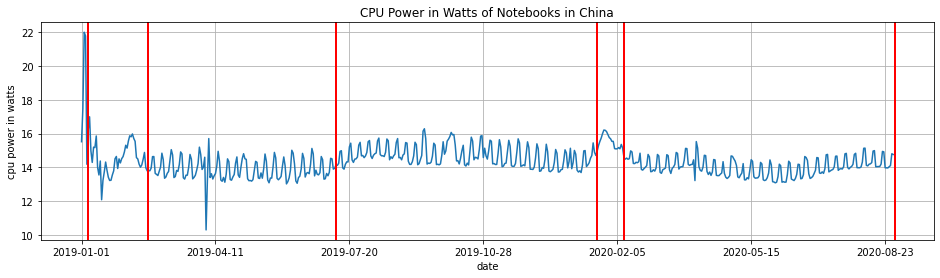}
    \caption{Changes in average CPU power in watts across all computers in China.}
    \label{fig:intensityEffect}
\end{figure}

After exploring the changes in the quantity of usage, we turned our attention to the intensity of usage, focusing on the average CPU power in watts across all computers in China. As depicted in Figure~\ref{fig:intensityEffect}, there was a noticeable increase in average CPU power around February 1st, 2020, which closely follows the implementation of China's Workplace Closing policy on January 26th, 2020. The algorithmic analysis also identified a change point around the onset of the COVID-19 pandemic, suggesting its impact on CPU power.

\begin{figure}[htbp]
    \centering
    \begin{subfigure}{0.7\textwidth}
        \includegraphics[width=\textwidth]{figs/image31.png}
        \caption{Notebooks}
        \label{fig:CPU-Notebooks}
    \end{subfigure}
    
    \begin{subfigure}{0.7\textwidth}
        \includegraphics[width=\textwidth]{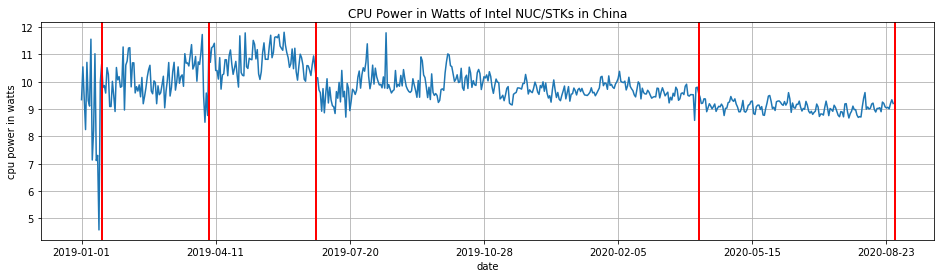}
        \caption{Desktops}
        \label{fig:CPU-Desktops}
    \end{subfigure}
    
    \begin{subfigure}{0.7\textwidth}
        \includegraphics[width=\textwidth]{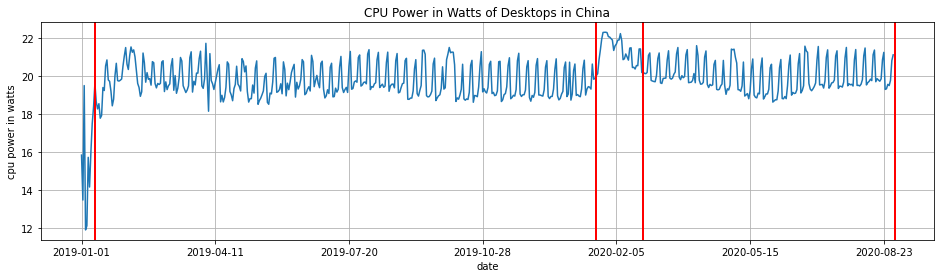}
        \caption{Intel NUC/STK}
        \label{fig:CPU-NUC}
    \end{subfigure}
    
    \caption{Impact of Workplace Closing Policy on CPU power across different types of computers.}
    \label{fig:CPUtypes}
\end{figure}

To understand the impact of the Workplace Closing Policy on various types of computers, we segmented the dataset by chassis types. Figure~\ref{fig:CPUtypes} illustrates the changes in CPU power across Notebooks, Desktops, and Intel NUC/STK. Unlike Notebooks and Desktops, Intel NUC/STK did not see an uptick in CPU power at the onset of the pandemic. For Notebooks, we observed an increase of approximately 1 watt in the average CPU power around the time of the initial activation of the Workplace Closing Policy. This increased usage reverted to baseline levels after roughly 20 days. Desktops experienced a larger increase of about 2 watts, which normalized after about one month. The red lines in the graphs denote breakpoints identified through the method outlined in Section 3.2.3. These breakpoints for Notebooks and Desktops align closely with the time of policy implementation. On the contrary, Intel NUC/STK did not display a significant shift in CPU power, possibly because these devices lack a direct user interface (these tend to be used as mini file servers or to drive screens in public spaces). Overall, our findings suggest an increase in the intensity of computer usage following the activation of the policy, albeit not uniformly across all computer types.  For example, we know that gaming is typically more intensive and leads to higher computer power.  We will test this hypothesis using the “persona” information contained in DCA in the following section.

\subsubsection{Activation of Workplace Closing Policy on the Number of Different Types of Users
}

\begin{figure}
    \centering
    \includegraphics[width=.95\textwidth]{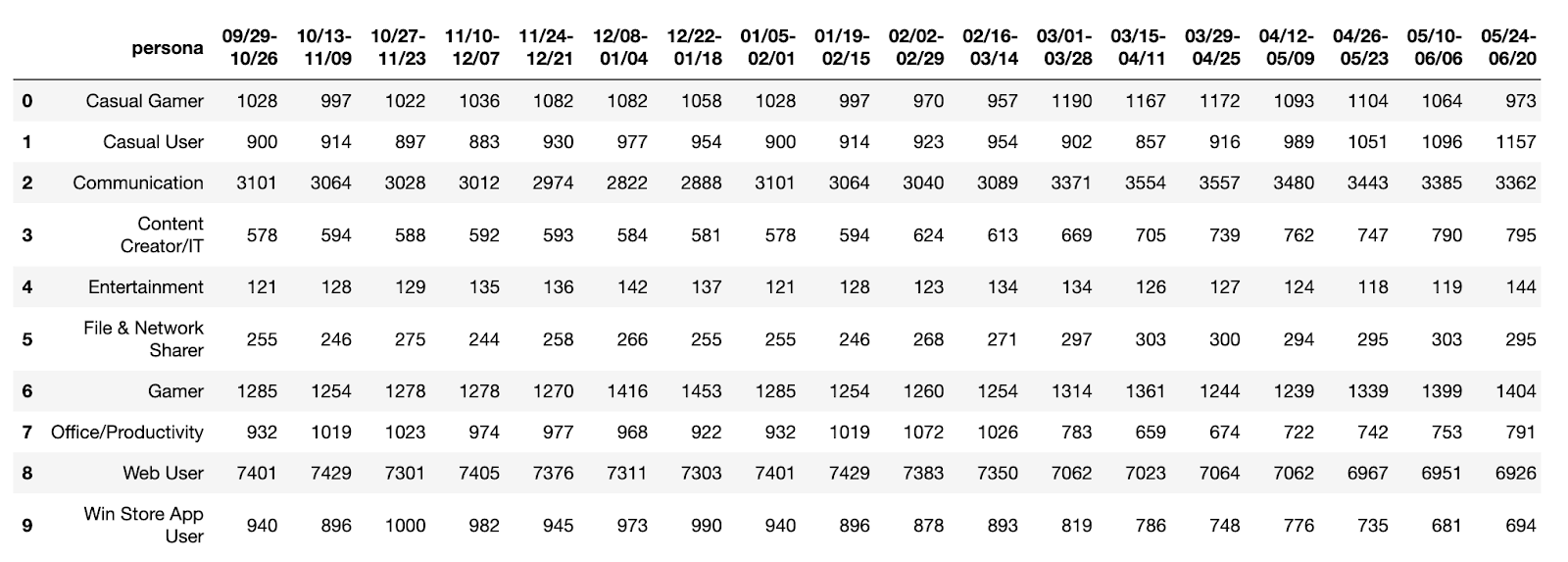}
    \caption{Number of each type of user in 4-week intervals with a 2-week sliding window}\label{tab:PersonaCounts}
\end{figure}

In addition to national-level analysis, we extended our focus to individual user behavior. Our DCA dataset contains a "persona" label for each user, initially determined via a k-means clustering algorithm as of March 14th, 2019. To monitor shifts in behavior during the pandemic, we reapplied this algorithm to current app usage data, averaged over four-week periods and analyzed through a two-week sliding window. This approach ensured a smoother trend line while maintaining the original centroids for comparison of user groups before and after the onset of COVID-19. Figure ~\ref{tab:PersonaCounts} presents the counts for each persona type at four-week intervals, calculated using a two-week sliding window, covering the period from September 2019 to June 2020.

\begin{figure}
    \centering
    \includegraphics[width=.8\textwidth]{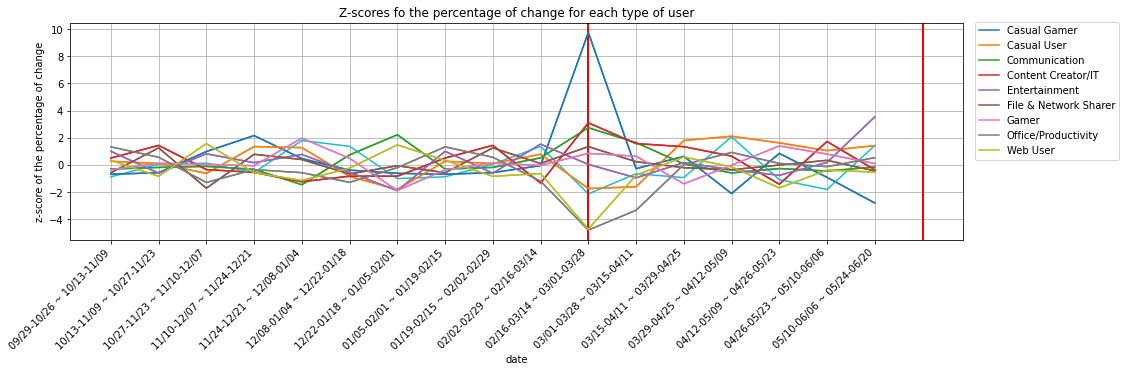}
    \caption{Changes in normalized categories across windows.}
    \label{fig:PersonaZScores}
\end{figure}

The population of user personas showed significant variation, ranging from 250 in the "File \& Network Sharer" category to 7000 for "Web Users." To standardize these numbers for comparison, we calculated both the differences between two consecutive four-week intervals and the z-scores for each persona. Figure~\ref{fig:PersonaZScores} illustrates the normalized changes in user counts, revealing a notable shift around late March 2020, immediately following the activation of the Workplace Closing Policy.  More specifically, an algorithmic analysis of the change points reveals what is obvious from the figure, there is a significant change point that occurs between the window of mid-February to mid-March 2020 and the window of beginning March to end of March 2020. This is the first point at which the earlier window does not contain the activation of the Workplace Closing Policy and the latter window does contain this activation.

More specifically, the "Casual Gamers" category showed the largest increase, while the "Communication Users"  (e.g. Zoom, Teams, etc.) and "Content Creators" categories also grew. Conversely, the number of "Web" and "Office/Productivity" users declined, suggesting that the policy's activation had a differential impact on various user categories.

\section{Conclusion}
Our study elucidates a causal link between the activation of the Workplace Closing Policy and subsequent changes in computer usage patterns, both at an aggregate and individual level. We found that following the policy activation, computer usage for notebooks, desktops, and 2-in-1s increased between 1.2 to 2.3 hours, with high-end CPUs (i7 and i9) showing greater increments than low-end CPUs (i5). Specifically, 2-in-1s experienced the most significant surge in usage, followed by notebooks and desktops. Similarly, CPU powerin notebooks and desktops increased temporarily before reverting to baseline levels. On the user persona front, the policy led to an uptick in the number of gaming, content creation, and communication users, while office and productivity-focused users declined.

The policy's deactivation yielded less pronounced effects, with a marginal decrease of 0.1 to 0.5 hours in computer usage across various CPU categories. Interestingly, the decrease was not consistently higher for high-end CPUs compared to their low-end counterparts. 
For future work, we aim to investigate how other types of policies, such as transit restrictions and stay-at-home orders, impact computer usage both during activation and deactivation of such policies. 

\section{Data Availability}
The Oxford COVID-19 tracking data that supports the findings of this study was collected by Oxford University, and is publicly available on their Github page: \\
\url{https://github.com/OxCGRT/covid-policy-dataset}.

The Intel DCA data that supports the findings of this study are available from Intel Corporation, but restrictions apply to the availability of this data. The data was used under license and agreement for the current study, and so the data is not publicly available. The data can be made available from Bijan Arbab upon reasonable request and with permission of Intel Corporation.

\bibliographystyle{plain} 
\bibliography{sample}

\end{document}